\documentstyle[prb,preprint,aps]{revtex}
\begin{document}
\draft
\preprint{}
\title{Self-consistent calculation of ionized impurity
scattering in semiconductor quantum wires}
\author{Ben Yu-Kuang Hu\cite{byline} and S. Das Sarma}
\address{Department of Physics, University of Maryland,
   College Park, Maryland 20742-4111}
\date{May 28, 1993}
\maketitle
\medskip

\begin{abstract}
We calculate the electron elastic mean free path due to ionized
impurity scattering in semiconductor quantum wires, using
a scheme in which the screened ionized impurity potential and the
electron screening self-consistently determine each other.
By using a short-range scattering potential
model, we obtain an {\sl exact}
solution of the self-energy within the self-consistent Born
and ``noncrossing" approximations.
We find that, compared to the mean free path for the bare unscreened
potential
$\ell_{\rm bare}$, the calculated mean free path including
self-consistent
screening $\ell_{\rm sc}$ is {\sl substantially larger},
going as $\ell_{\rm sc} \sim \ell_{\rm bare}[\ln(\ell_{\rm bare})]^2$
for large $\ell_{\rm bare}$.
\end{abstract}


Advances in nanofabrication technology
now make it possible to fabricate semiconductor quantum wires
in which the motion of the electrons is confined to be in
one spatial direction.  Quantum wires have the potential
for many microelectronics applications, such as novel
optoelectronic devices and transistors, and many experiments probing the
electronic properties of these quantum wires have been performed.\cite{exper}
It is, of course, experimentally desirable to make these wires as clean
as possible.  However, to populate the wires with carriers, it is necessary
to place ionized dopants close to the wire, and these dopants scatter the
free carriers in the wire.  Thus, elastic impurity scattering cannot
be avoided in conducting quantum wires. A quantitative understanding of
ionized impurity scattering in quantum wires is thus important from both
technological and fundamental viewpoints.

Sakaki\cite{sakaki} made the striking prediction that in quantum wires
that are remotely doped, the electrons should have extremely high mobilities
because elastic ionized impurity scattering for Fermi surface electrons
is greatly suppressed.   This is due to the fact that at low
temperatures, the only allowed resistive scattering in a one-dimensional
system is the $2 k_{\rm F}$ scattering.
However, in Sakaki's calculation, and in some subsequent calculations by
others,\cite{johnson} the interaction between
the ionized donors and the conduction electrons
was assumed to be the {\sl bare} unscreened Coulomb potential.  This
ignores screening effects of the electrons in the
quantum wire.  In this paper, we show, using many-body techniques, that
the inclusion of screening effects in the wire {\sl substantially enhances}
the mean free path of electrons at the Fermi surface.
The mean free path for electrons at the Fermi surface is particularly
important because this is the quantity which determines
the low field transport properties in these structures.  Furthermore,
it has been shown that the mean free path and the localization
length of electrons in strictly one-dimensional systems are essentially
equivalent,\cite{leerama} and thus the mean free path of electrons
is the maximum length over which a quantum wire can be considered metallic.
This gives added impetus to an accurate calculation of the electronic
mean free path in quantum wires.

The simplest approximation taking screening into account
is one which assumes that disorder has no effect on the screening
properties of the one-dimensional electron gas.\cite{fishman}
Unfortunately, in the $T=0$ limit, this approximation gives meaningless
results for the mean free path of electrons at the Fermi surface,
for the following reason.
At the Fermi surface, the only elastic scattering that can occur is
the transfer of electrons from one side of the Fermi surface to
the other, with momentum transfer of $2 k_F$.  To the lowest order,
this depends
on the $2 k_F$ component of the screened impurity
potential.  However, as is well known, the $2k_{\rm F}$ screening at $T=0$
in a pure one-dimensional wire is {\sl perfect} ({\it i.e.}, the $2 k_{\rm
F}$ screening diverges), and hence
within this approximation, the scattering rate at the Fermi surface
is always zero for any impurity concentration. Clearly one needs
a better (and more physical) approximation.

The key point is that disorder affects screening
which will no longer be perfect at $2k_F$ in the presence of impurities.
The extent to which screening in the presence of scattering
is modified from the pure wire case
depends on the strength of the disorder created by the ionized impurities,
but the strength of disorder in turn depends on the strength of screening.
Thus, we are faced with a problem in which both the scattering
potential and the screening properties self-consistently determine
each other.  In this paper, we carry out this self-consistent scheme
in one-dimensional wires,
which has been previously applied to two-dimensional systems in
a large magnetic field,\cite{xie} but to the best of our knowledge, has not
yet been applied to one-dimensional systems.

In our model, we assume that (1) the electrons are confined to a
one-dimensional quantum wire in the strip geometry with
a finite width $w$ in infinite square well confinement, and
(2) the ionized donors are placed randomly along a line
a distance $d$ away from the wire.
The set of equations which must be solved for this calculation
is shown schematically in Fig.\ 1.
For a given screened impurity potential, the Green's function
is computed.  In this paper, we employ both the Born approximation and
what we denote the ``noncrossing approximation," which
excludes all diagrams that have impurity lines crossing one another.
(In fact, as we show below, for experimentally relevant parameters we find
that Born and noncrossing approximations give essentially
indistinguishable results.) Then, with these Green's
function, one computes the polarizability, including the ladder vertex
corrections (which are necessary to satisfy the Ward identities and
ensure, {\it inter alia}, that particles are conserved).
This gives the dielectric function,\cite{lai} which is used to
screen the disordered impurity potential, yielding a new screened
impurity potential.  This loop is iterated until convergence is complete.

The actual bare electron--impurity interaction should of course be the
long-ranged Coulomb interaction, and ideally, the exact interaction
should be used in the calculation,
but the numerical task in performing the iteration
to self-consistency is formidable.  Therefore, as a first attempt, we
find it expedient to approximate the screened interaction to be of
the short-ranged form $U_0\,\delta(x-x_0)$,
({\it i.e.}, a constant in momentum space).
This approximation is valid because the actual screened interaction
in quantum wires is short-ranged due to screening by
the conduction electrons.  Our primary motivation in using the short-range
scattering potential model is that we are able to obtain physically
meaningful analytical results whereas the Coulomb case is necessarily
completely numerical in the self-consistent scheme.

The short-ranged impurity potential approximation yields simple
expressions, both in the Born and noncrossing approximations, for the
self-energy $\Sigma$ and the static random phase approximation (RPA)
polarizability $\Pi(q)$ with vertex corrections.\cite{lai,dassarma}
For both approximations, the self-energy is $k$-independent
and is given, for arbitrary temperature, by [in this and subsequent
equations, to obtain the result for the Born (noncrossing) approximation,
the terms within $[\![\cdots ]\!]$ should be excluded (included)]
\begin{eqnarray}
\Sigma(\omega)
=&& {N_{\rm i}U_0^2}\int_{-\infty}^\infty {dq\over 2\pi}\, G(q,\omega)\ \
\biggl[\!\!\!\;\biggl[
+ N_{\rm i}U_0 \sum_{n=2}^\infty  \Bigl\{U_0\,\int_{-\infty}^\infty
{dq\over 2\pi} G(q,\omega)\Bigr\}^n \biggr]\!\!\!\;\biggr],\nonumber\\
G(q,\omega) = &&{1\over \hbar\omega - \xi_q - \Sigma(\omega)}.
\label{eq1}
\end{eqnarray}
Here $N_{\rm i}$ is the impurity concentration, $m$ is the band
electron mass,
and $\xi_q = \hbar^2q^2/(2m) - \mu$, where
$\mu$ is the chemical potential (which is computed self-consistently).
This implies that $\Sigma(z)$ can be obtained from the
following cubic equation
\begin{eqnarray}
\Bigl(x^2+\mu+\hbar \omega\Bigr) \Bigl( x \Bigl[\!\!\!\:\Bigl[
- {U_0 k_{\rm F}\over 2}
\Bigr]\!\!\!\:\Bigr]\Bigr) - {\hbar\gamma\over 2} =&& 0,\nonumber\\
\Sigma = x^2 + \hbar\omega + \mu,&&
\label{eq2}
\end{eqnarray}
where $\gamma = 2 N_i U_0^2 m/(\hbar^3 k_F)$ is the Born
approximation scattering rate for electrons at the Fermi surface.
The proper choice for the correct solution from the three roots
of Eq.\ (\ref{eq2}) is dictated by the requirement that
${\rm Im}[\Sigma(\omega)]  { {\smash{\lower2pt\hbox{$\scriptstyle<$}}} \atop
{\smash{\raise2pt\hbox{$\scriptstyle>$}} } } \, 0$ for
${\rm Im}[\omega]  { {\smash{\lower2pt\hbox{$\scriptstyle>$}}} \atop
{\smash{\raise2pt\hbox{$\scriptstyle<$}} } } \, 0$.   In Fig.\ 2,
we show the self-energy for both approximations
with $\hbar\gamma/E_{\rm F} = 0.5.$  Note that the magnitude of
${\rm Im}[\Sigma(\omega)]$ is {\sl smaller} in the noncrossing approximation
than in the Born approximation, because the Born approximation
overestimates the scattering of a particle from a $\delta$-function
potential.

The static RPA polarizability with the ladder
vertex corrections (Fig.\ 1) in a one-dimensional electron gas is
found to be
\begin{eqnarray}
\Pi(q) =&& 2 k_{\rm F} k_{\rm B} T \sum_{i\nu_n}
\biggl(E_{\rm F}^{1/2}\sqrt{\Sigma(i\nu_n)-i\hbar\nu_n-\mu}\
\Bigl({\hbar^2q^2\over 2m} +
4\{\Sigma(i\nu_n) - i\hbar\nu_n - \mu\}\Bigr) \nonumber\\
&&\qquad\qquad - \hbar\gamma E_{\rm F}
\Bigl[\!\!\!\:\Bigl[
\times \{ 1 + N_i^{-1}U_0^{-1}\Sigma(i\nu_n)\}
\Bigr]\!\!\!\:\Bigr]\biggr)^{-1},
\label{eq3}
\end{eqnarray}
where the summation is over the
``fermion frequencies" $i\hbar\nu_n = (2n+1)\pi k_{\rm B}T$.
While the above equations are valid for finite temperature,
henceforth, we specialize to the case of $T=0$, where the effects
of self-consistent screening are most pronounced.
In Fig.\ 3, we show the calculated
$\Pi(q)$ for various values of the strength of disorder in both
the Born and noncrossing approximations.
The suppression of the Kohn anomaly in $\Pi(q=2k_{\rm F})$ is larger
in the Born approximation than in the noncrossing approximation,
which is consistent with the fact that the Born approximation
overestimates scattering.

We incorporate the $\delta$-function screened potential approximation
into our iteration scheme by adjusting the potential strength $U_0$ at
each iteration to match the Born approximation scattering rate
at the Fermi surface.  The self-consistent scattering rate
$\gamma_{\rm sc}$ is calculated as follows.
For a set of parameters (doping density, etc.), we pick an initial $U_0$
to match the scattering rate of the bare Coulomb interaction at the
Fermi surface.  The polarizability is then calculated for that $U_0$
(for both the separate cases of the Born and noncrossing approximations).
This allows us to recalculate the scattering rate for the screened Coulomb
potential at the Fermi surface.  The $U_0$ is then adjusted to
reproduce this scattering rate, and the polarizability is recomputed.
This procedure is repeated until convergence is complete, which
typically takes 10 -- 50 number of iterations.
The chemical potential is also adjusted to ensure that the density of
electrons is constant, as shown in the inset of Fig.\ 3.

The results are shown in the Fig.\ 4, for the wire density $n = 10^6\,{\rm
cm}^{-1}$ and wire width $w=100\,{\rm \AA}$.  We show the mean free path
$\ell=v_F/\gamma$, for both the bare and the self-consistently screened
Coulomb interactions with both approximations,
as a function of the distance $d$ of the impurities
from the wire.  The mean free path for the bare Coulomb interaction,
$\ell_{\rm bare}$ is a reproduction of Sakaki's result,\cite{sakaki}
generalized to take into account the finite width of the wire.
As can be seen from Fig.\ 4, both Born and
noncrossing approximations give essentially indistinguishable results
for self-consistently screened mean free paths $\ell_{\rm sc}$,
indicating that multiple scatterings from the same impurity do not play
an important r\^ole in the transport properties of these systems.
Also note that the self-consistently calculated mean free paths
are significantly longer than the unscreened mean free paths.
The large enhancement is a result of the Kohn anomaly in $\Pi(q=2k_{\rm F})$
(see Fig.\ 3) which diverges logarithmically as the scattering
goes to zero. Thus, the self-consistent scattering rate $\gamma_{\rm
sc}$, in the limit where the bare scattering rate $\gamma_{\rm bare}$ is
small, goes as $\gamma_{\rm sc} =\gamma_{\rm bare}/|\epsilon(2 k_F)|^2
\sim \gamma_{\rm bare}/ |\ln(\gamma_{\rm sc})|^2$,
which implies that the self-consistent mean free path should
go (to lowest order) as $\ell_{\rm sc} \sim \ell_{\rm bare}\,|\ln(\ell_{\rm
bare})|^2$.  We have explicitly verified this asymptotic behavior in
our numerical calculation.

This result seems to indicate that semiconductor quantum wires
will be very good conductors, and will be metallic over very large
length scales, because the elastic mean free path is essentially a
measure of the localization length in one-dimensional systems.
Experimentally, however, the opposite seems to be
true.  Attempts to fabricate quantum wires using electrostatic
confinement with split-gate generally fail if the length of the wire
is more than a few microns,\cite{behringer} and it has been
theoretically shown\cite{nixon} that these wires tend to become
disjointed before single channel occupation is achieved.   The
reason for this extreme sensitivity to impurities in the electrostatically
confined wires is that the confinement potential is very
shallow,\cite{davies} and therefore perturbations due to ionized
impurities have a very large effect.
However, recently, wires of GaAs in AlGaAs have
been fabricated,\cite{kapon} where the confinement potential
is much stronger, and these wires are more robust against the effects
of the ionized dopant potential.
These wires should show dramatically long mean free paths at the Fermi
surface, as calculated in this paper.  Our theory is only applicable to
the situation with weak disorder, where $\hbar\gamma/E_{\rm F} \lesssim 1$.

Before concluding, we discuss the validity of our assumptions and
approximations, and possible extensions of this calculation.
It is natural to ask if multiple scattering diagrams we have left out of our
self-consistent iteration scheme, such as the one shown in
Fig.\ 1(b), are important.   We have evaluated this diagram
explicitly, assuming a $\delta$-function interaction,
and a $k$-independent self-energy in the Green's function,
and we obtain
\begin{equation}
\Sigma_2(k,\omega) = {3 E_{\rm F}\ \over 4}
{(\hbar \gamma)^2\over
\Bigl[\hbar\omega+\mu-\Sigma(\omega)\Bigr]
\Bigl[\hbar^2 k^2/(2m) - 9(\hbar\omega+\mu-\Sigma(\omega))\Bigr]}.
\label{eq4}
\end{equation}
Comparing Eqs.\ (\ref{eq2}) and (\ref{eq4}), one can see that
$\Sigma_2$ is smaller by a factor of $\hbar\gamma/E_{\rm F}$ than
the Born and noncrossing approximation $\Sigma$'s, and therefore
since $\hbar\gamma/E_{\rm F} < 0.1$ in our self-consistent calculation,
$\Sigma_2$ makes a negligible contribution.
We find that the results for the calculated
self-consistently screened mean free paths are relatively insensitive
to the criterion used for the adjustment of $U_0$ to match the
actual screened Coulomb interaction.
However, it is probable that the scheme we have used
somewhat overestimates the effect of self-consistent screening,
since we are looking at the
scattering rates where the effect of screening is the strongest ({\it
i.e.}, at the Fermi surface).  On the other hand,
we have neglected the effect of electron--electron vertex corrections
in the polarizability, which have been shown to {\sl increase} the
divergence at $2k_F$.\cite{luther}  To extract the true scattering rate,
one must use the actual Coulomb interaction for
the bare electron--ionized-impurity interaction.  This considerably
complicates the iteration task, due to the lack of any analytic results
and the strong $k$ dependence of the self-energy, and work on this is
currently in progress.  We expect our short-ranged results to be at least
semiquantitatively valid.  The issue of the applicability of Fermi
liquid theory to disordered interacting one-dimensional systems
is complicated, and has been discussed previously.\cite{flt}
Finally, we note that our calculation can easily be extended to
include effects of finite temperature and occupancy of several subbands.
The finite temperature will serve to cut off the $2k_{\rm F}$ divergence
of screening, and therefore the maximum polarizability at $2k_F$
will be on the order of $\ln[{\rm max}\{k_{\rm B}T,\hbar\gamma\}]$.

To conclude, using a self-consistent screening scheme,
we have calculated the elastic mean free path of electrons in a
quantum wire which are scattered by remote ionized impurities.
We have shown that the mean free paths of electrons at the Fermi surface
is significantly enhanced by the self-consistent screening.
The calculated mean free paths provide a direct measure of the length scale
over which the electrons in quantum wires are expected to exhibit metallic
behavior due to the peculiarity of Anderson localization in
one-dimensional disordered systems.

This work is supported by the U.S. ONR and the U.S. ARO.

\begin{figure}
\caption{(a) Set of diagrams which are solved
to calculate of the electron Green's function
in the self-consistent Born approximation (excluding diagrams in parentheses)
and noncrossing approximation (including diagrams in parentheses).
The thin (thick) dashed [wavy] lines are the bare (dressed)
electron--impurity [electron--electron] interactions.
The screened electron--impurity interaction and the Green's function
are numerically iterated until self-consistency is obtained.
(b) Second-order contribution to self-energy, which we ignore in
self-consistent loop.}
\label{fig1}
\end{figure}

\begin{figure}
\caption{Real (solid lines) and imaginary (broken lines) parts of
the electron self-energies for impurity scattering, as a function of
frequency, for $\hbar\gamma/E_{\rm F} = 0.5$,
in the self-consistent Born (thick bold lines) and noncrossing
(medium bold lines) approximations.  For comparison, we also show
the $\Sigma$ for the non-self-consistent Born approximation (thin
lines), in which the bare (instead of dressed) Green's function is used.}
\label{fig2}
\end{figure}

\begin{figure}
\caption{Static one-dimensional polarizabilities, $\Pi(q)$, within the
RPA approximation with short-ranged impurities (including vertex
corrections) at $T=0$, for $\hbar\gamma/E_{\rm F} = 0,0.1,0.5$.
The bold lines are for the self-consistent Born approximation
and the thin lines are for the noncrossing approximation.
There is a larger suppression of the $2 k_F$ peak in the Born
approximation as compared to the noncrossing approximation, due to
the fact that the Born approximation overestimates the scattering.
The inset shows the chemical potential as a function of disorder
in both approximations.}
\label{fig3}
\end{figure}

\begin{figure}
\caption{Elastic mean free paths $\ell$ as a function of distance of charged
impurities $d$ from a quantum wire.  The parameters used were for GaAs,
with electron density of $10^6\,{\rm cm}^{-1}$ and wire width of
$100\,{\rm\AA}$.  The solid line is for the
self-consistently screened potentials, and the broken line is for the
bare Coulomb interaction. On this plot, the result for
noncrossing approximation is essentially indistinguishable
from that of the self-consistent Born approximation.
In the inset, we show the ratio of the self-consistently
screened to bare mean free paths, as a function of $d$.}
\label{fig4}
\end{figure}

\end{document}